\documentstyle[12pt,moriond,epsfig,axodraw]{article}

\def\be{\begin{equation}}
\def\ee{\end{equation}}
\def\bc{\begin{center}}
\def\ec{\end{center}}
\def\bea{\begin{eqnarray}}
\def\eea{\end{eqnarray}}

\def\fz{{\psi_z}}

\def\lsq{{\Lambda^2}}
\def\mgau{M_{\lambda}}

\def\muc{{\mu^c}}

\def\simlt{\stackrel{<}{{}_\sim}}

\def\tm{\tilde{\mu}}
\def\tmc{{\tilde{\mu^c}}}

\catcode`@=11
\def\marginnote#1{}
\newcount\hour
\newcount\minute
\newtoks\amorpm
\hour=\time\divide\hour by60
\minute=\time{\multiply\hour by60 \global\advance\minute by-\hour}
\edef\standardtime{{\ifnum\hour<12 \global\amorpm={am}%
        \else\global\amorpm={pm}\advance\hour by-12 \fi
        \ifnum\hour=0 \hour=12 \fi
        \number\hour:\ifnum\minute<10 0\fi\number\minute\the\amorpm}}
\edef\militarytime{\number\hour:\ifnum\minute<10 0\fi\number\minute}
\def\draftlabel#1{{\@bsphack\if@filesw {\let\thepage\relax
   \xdef\@gtempa{\write\@auxout{\string
      \newlabel{#1}{{\@currentlabel}{\thepage}}}}}\@gtempa
   \if@nobreak \ifvmode\nobreak\fi\fi\fi\@esphack}
        \gdef\@eqnlabel{#1}}
\def\@eqnlabel{}
\def\@vacuum{}
\def\draftmarginnote#1{\marginpar{\raggedright\scriptsize\tt#1}}
\def\draft{\oddsidemargin 0.0truein
        \def\@oddfoot{\sl preliminary draft \hfil
        \rm\thepage\hfil\sl\today\quad\militarytime}
        \let\@evenfoot\@oddfoot \overfullrule 3pt
        \let\label=\draftlabel
        \let\marginnote=\draftmarginnote
   \def\@eqnnum{(\theequation)\rlap{\kern\marginparsep\tt\@eqnlabel}%
\global\let\@eqnlabel\@vacuum}  }
\catcode`@=12
\begin{document}
\vspace*{1cm}
\title{ Muon (g-2) in models with a very light gravitino}

\author{ELENA PERAZZI}

\address{Istituto Nazionale di Fisica Nucleare, Sezione di Padova,\\
Dipartimento di Fisica `G.~Galilei', Universit\`a di Padova, 
\\
Via Marzolo n.8, I-35131 Padua, Italy}

\maketitle
\abstracts{We present the results of a general analysis of the
contributions of the supersymmetry-breaking sector to the
muon anomalous magnetic moment in models with a superlight gravitino.
We find constraints on the model parameters that are comparable and 
complementary to the ones coming from collider searches, and will
become more stringent after the Brookhaven E821 experiment.}

\vskip2truecm
The measurement of the anomalous magnetic moment of the muon, 
\be
\label{a1}
a_{\mu}^{ex} \equiv (g_{\mu} - 2)/2 =  (116592350  \pm  730)  \times  10^{-11} \, ,
\ee
together with the theoretical prediction of the Standard Model,
\be
\label{a2}
a_{\mu}^{SM} =  (116591596  \pm   67)  \times  10^{-11} \, , 
\ee
provides a very powerful constraint on extensions of the SM:  
from comparison between eqs.~(\ref{a1}) and ~(\ref{a2}) 
we find, for  possible contributions $\delta a_{\mu}$ from New Physics,  
the $95 \%$ confidence level bound: 
$- 0.7 \times 10^{-8} < \delta a_{\mu} < 2.2 \times 10^{-8}$. 
Moreover, the E821 experiment at Brookhaven ~\cite{e821}
is expected to 
reduce further the experimental error by roughly a factor of 20.

The full 
one-loop contribution to $a_{\mu}$ in the Minimal 
Supersymmetric Standard Model (MSSM) 
is well known: 
the only case in which 
$\delta a_{\mu}$ can become relevant
is when the masses of supersymmetric particles are close to 
their present lower bounds and $\tan \beta = v_2/v_1$ is very
large.

We performed~\cite{bpz}a general analysis of the one-loop contributions
to $\delta a_{\mu}$ in supersymmetric models with a light gravitino, 
where the effective 
low-energy theory contains, besides the MSSM states,  
also the gravitino and its superpartners.
Most of the 
existing calculations of these effects were performed in the 
framework of supergravity (the most complete one is that of  ref.~\cite{paco}); however, 
in the case of a
light gravitino (the only phenomenologically relevant one for this 
type of study), we can work directly in the globally supersymmetric 
limit, keeping only the goldstino and its spin--0 superpartners 
(sgoldstinos) as the relevant degrees of freedom from the 
supersymmetry-breaking sector.

We constructed~\cite{bpz}the most general
$N=1$ globally supersymmetric model whose physical content consists of:
$U(1)$ 
gauge boson and gaugino $(A_{\mu},\lambda)$,
associated 
with the exact gauge symmetry of supersymmetric QED; 
left-handed muon and related smuon $(\mu,\tm)$ (with charge $-1$);
 left-handed anti-muon and related smuon $(\muc,\tmc)$ (with charge $+1$);
finally, neutral goldstino and sgoldstino
$(\fz,z)$.

The model is characterized by the following spectrum
parameters: the muon mass $m_{\mu}$;
two different squared masses $m_{1}^{2}$ and $m_{2}^{2}$ 
and a mixing angle $\theta$ (related to the smuon off-diagonal mass parameter
$\delta m^{2}\equiv( m_1^{2}-m_2^{2})\  sin(2 \theta)/2\ $) for the smuons;  
two different squared masses 
$m_{S}^{2}$ and $m_{P}^{2}$ for the 
sgoldstinos 
($z \equiv (S + i P)/\sqrt{2}$); the photino mass $M_{\lambda}$.

In addition, we must consider the supersymmetry-breaking scale, $\sqrt{F}$, and
two additional parameters, $\gamma_f$ and $\gamma_K$.
These are all the parameters relevant for the one-loop computation of $\delta a_{\mu}$.

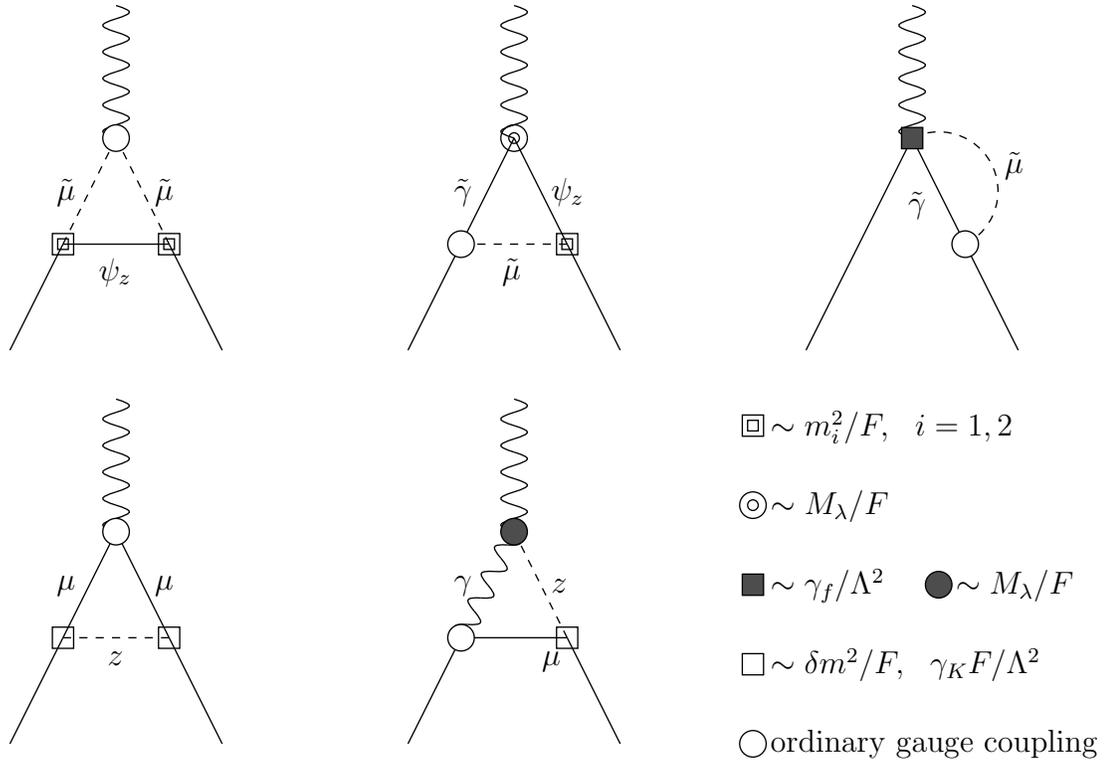
\begin{figure}[htb]
\begin{center}
\begin{picture}(450,170)(30,0)
\Photon(100,170)(100,120){5}{5}
\DashLine(80,80)(100,120){3}
\Line(60,40)(80,80)
\DashLine(100,120)(120,80){3}
\Line(120,80)(140,40)
\Line(80,80)(120,80)
\Text(85,100)[r]{$\tilde{\mu}$}
\Text(115,100)[l]{$\tilde{\mu}$}
\Text(100,75)[t]{$\psi_z$}
\BCirc(100, 120){5}
\EBox(78,78)(82,82)
\EBox(76,76)(84,84)
\EBox(118,78)(122,82)
\EBox(116,76)(124,84)
%
\Photon(250,170)(250,120){5}{5}
\Line(230,80)(250,120)
\Line(210,40)(230,80)
\Line(250,120)(270,80)
\Line(270,80)(290,40)
\DashLine(230,80)(270,80){3}
\Text(235,100)[r]{$\tilde{\gamma}$}
\Text(265,100)[l]{$\psi_{z}$}
\Text(250,75)[t]{$\tilde{\mu}$}
\BCirc(230,80){5}
\CArc(250,120)(5,0,180)
\CArc(250,120)(5,180,0)
\CArc(250,120)(2,0,180)
\CArc(250,120)(2,180,0)
\EBox(266,76)(274,84)
\EBox(268,78)(272,82)
\Photon(400,170)(400,120){5}{5}
\Line(360,40)(400,120)
\Line(400,120)(440,40)
\DashCArc(410,100)(22.36,-63.435,116.565){3}
\Text(400,95)[l]{$\tilde{\gamma}$}
\Text(440,115)[t]{$\tilde{\mu}$}
\GBox(396,116)(404,124){0.3}
\BCirc(420,80){5}
\end{picture}
\end{center}
\begin{center}
\begin{picture}(450,170)(30,-25)
\Photon(100,170)(100,120){5}{5}
\Line(80,80)(100,120)
\Line(60,40)(80,80)
\Line(100,120)(120,80)
\Line(120,80)(140,40)
\DashLine(80,80)(120,80){3}
\Text(100,75)[t]{$z$}
\Text(85,100)[r]{$\mu$}
\Text(115,100)[l]{$\mu$}
\BCirc(100,120){5}
\EBox(76,76)(84,84)
\EBox(116,76)(124,84)

\Photon(250,170)(250,120){5}{5}
\Photon(230,80)(250,120){4}{4}
\Line(210,40)(230,80)
\DashLine(250,120)(270,80){3}
\Line(270,80)(290,40)
\Line(230,80)(270,80)
\Text(235,100)[r]{$\gamma$}
\Text(265,100)[l]{$z$}
\Text(265,75)[t]{$\mu$}
\GCirc(250,120){5}{0.3}
\BCirc(230,80){5}
\EBox(266,76)(274,84)
\BCirc(340,40){5}
\Text(348,40)[l]{ordinary gauge coupling}
\EBox(338,158)(342,162)
\EBox(336,156)(344,164)
\Text(348,160)[l]{$\sim m_{i}^{2}/F,\ \  i=1,2$}
\CArc(340,130)(5,0,180)
\CArc(340,130)(5,180,0)
\CArc(340,130)(2,0,180)
\CArc(340,130)(2,180,0)
\Text(348,130)[l]{$\sim M_{\lambda}/F$}
\GBox(336,96)(344,104){0.3}
\Text(348,100)[l]{$\sim \gamma_{f}/\Lambda^{2}$}
\EBox(336,66)(344,74)
\Text(348,70)[l]{$\sim \delta m^{2}/F,\ \ \gamma_K F/\Lambda^{2}$}
\GCirc(410,100){5}{0.3}
\Text(418,100)[l]{$\sim M_{\lambda}/F$}

\end{picture}
\end{center}
\vspace{-2cm}
\caption{\em The different classes of one-loop diagrams 
contributing to $a_{\mu}$, together with the parameters
controlling the different couplings}
\label{fig1}
\end{figure}

In the model described above, the different 
classes of one-loop Feynman diagrams that may contribute to $a_{\mu}$
(in addition to the well-known QED and SQED ones) 
are displayed in fig.~\ref{fig1}.
In ref.~\cite{bpz}  we presented the results
in their most general form; here we would like to point
out their main features and to comment on
their possible interpretation.
The first thing to say is that
we found a logaritmically divergent
contribution to $a_{\mu}$ 

\be
\label{divg}
\delta a_{\mu}^{(DIV)} =\left(2 \gamma_K  - \gamma_f \right)
{m_{\mu} \mgau \over 
8 \pi^2 \lsq}   
\log {\Lambda_{UV}^2 \over \mu^2} \, ,
\ee
where $\Lambda_{UV}$ is a cutoff in momentum space, $\Lambda$ is
the suppression scale of non-renormalizable terms in the Lagrangian
and $\mu$ is the renormalization scale.
This result is in agreement with those 
obtained within the supergravity 
formalism~\cite{paco}. 

Since the existence of such a contribution limits somewhat 
the predictive power of our effective theory, we could look
for a class of divergence-free models. We could just impose 
$\gamma_f=2\gamma_K$, but it would be better
to address the question in terms of symmetry arguments.
Indeed, we presented~\cite{bpz}an 
R-symmetry which forbids the divergent one-loop contributions, 
but unfortunately it forbids also gaugino mass $M_{\lambda}$,
thus making difficult the construction of a realistic model
with the full Standard Model gauge group. 

A milder requirement may be to ask for a symmetry that forces the 
divergent contribution, eq.~(\ref{divg}),  to 
be at least proportional to $m_{\mu}^2$.
An obvious candidate is a chiral $U(1)_S$,
under which the doublets $(\mu,\tm)$  and $(\muc,\tmc)$
have charge assignments whose sum is different from zero.
Such a symmetry
would be explicitly broken by a small parameter to allow for the muon
mass term. As a consequence, $\gamma_f$ and $\gamma_K$
would be suppressed by the same small parameter 
(and, also, we would have $\delta m^2 \equiv 
m_{\mu} A$, with $A={\cal O}(M_S)$, the order of
typical  supersymmetry-breaking
masses).

In such a framework, reasonable choices of the ultraviolet cutoff
would produce contributions to $a_{\mu}$ that are of the same
order of magnitude as the finite ones.
Still, we cannot use the latter to make precise predictions on 
$a_{\mu}$ (in the absence of a satisfactory microscopic
theory)  
but only to derive some `naturalness' 
constraints on the model parameters (if we disregard the possibility 
of miracolous - or at least not yet understood-  cancellations).  

To discuss the phenomenological impact of the finite contributions,
it is useful to parameterize all of them in a uniform fashion
\be
\label{param}
\delta a_{\mu} \equiv {m_{\mu}^2 M_x^2 \over 16 \pi^2 F^2} \, ,
\ee
where the square mass parameter $M_x^2$ 
can be positive or negative. The above form separates the 
dependence of $\delta a_{\mu}$ on the spectrum (through $M_x$) 
from that on the supersymmetry-breaking scale $\sqrt{F}$. 

\begin{figure}[ht]
\epsfig{figure=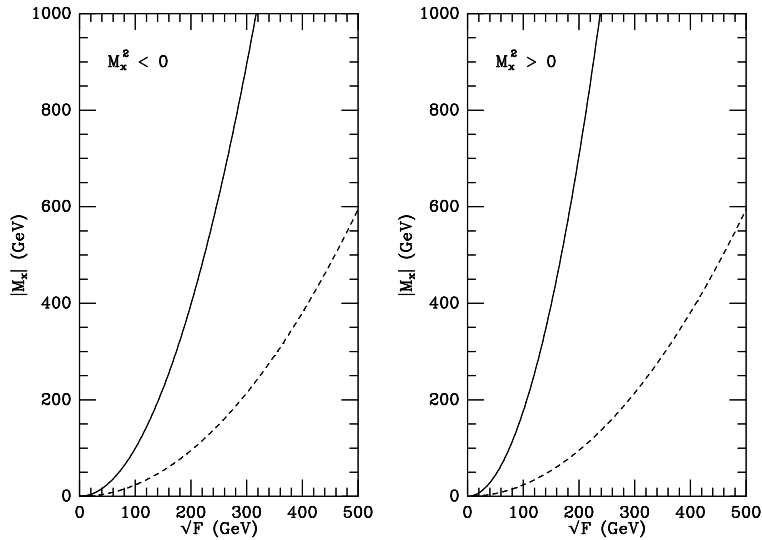,height=4cm,angle=-90}
\vspace{2cm}
\caption{{\it Contours of $\delta a_{\mu} \equiv m_{\mu}^2
M_x^2 / (16 \pi^2 F^2)$ in the plane $(\sqrt{F},|M_x|)$: the 
regions above the solid lines correspond to $\delta a_{\mu} 
< - 70 \; (> + 220) \times 10^{-10}$, and the dashed lines 
correspond to $\delta a_{\mu} = \mp 4 \times 10^{-10}$.}}
\label{ccc}
\end{figure}

The present experimental limit and the future sensitivity on $M_x$ 
are then shown in fig.~\ref{ccc}.
This information should be combined with 
the explicit expression of $M_x^2$ in terms of the 
spectrum, which can be easily read off the results of ref.~\cite{bpz}

In particular,
expressing 
$M_x^2$ as the sum of `goldstino' and `sgoldstino'
contributions, $M_x^2 = M^2_{x(G)}+M^2_{x(SG)}$, we have 
\be
\label{mxg}
M^2_{x(G)} = m_1^2 \left[ - {1 \over 6} + 
{\mgau^2 \over m_1^2 - \mgau^2} 
\left( 1- {\mgau^2 \over m_1^2 - \mgau^2} \log {m_1^2
\over \mgau^2} \right) \right] + (1 \to 2) \, ,
\ee
and 
\bea 
\label{mxsg}
M^2_{x(SG)} =
\left\{ \begin{array}{ll}
2 A \mgau \log {m_P^2 \over m_S^2} 
+ {\cal O}(A^2 m_{\mu}^2/m_{S,P}^2)
& (m_{\mu} \ll m_P,m_S)
\\ & \\
2 A \mgau  + A^2
&
(m_S,m_P \ll m_{\mu})
\end{array}
\right.
\, .
\eea

The size of the goldstino contribution to $a_{\mu}$ depends 
mainly on the smuon masses, and can be significant only for 
heavy smuons and a very low supersymmetry-breaking scale. 
The corresponding sign changes from positive to negative
for increasing smuon-to-photino mass ratio. 


The size of the sgoldstino contribution to $a_{\mu}$
depends on the photino mass, the sgoldstino masses
and the parameter $A\equiv\ \delta m^{2}/m_{\mu}$. 
The latter parameter plays a crucial role. Even if it would be
in contrast with the above-quoted chiral symmetry, $A$ could 
become  much heavier than the typical superparticle mass $M_S$
(e.g. it could be $\delta m^{2}={\cal O}(M_S^{2})$), and
the sgoldstino contribution could have an enhancement,
in particular in the extreme case of superlight sgoldstinos
$m_S, m_P
 \ll m_{\mu}$ (but
the qualitative picture remains the same whenever at 
least one of the sgoldstino masses is $\simlt m_{\mu}$).

%

In summary, with  the present  limits from accelerator searches 
(in particular the lower limit of roughly 200 GeV for $\sqrt{F}$), 
we can see that $a_{\mu}$  
provides a non-negligible but mild constraint. Such a constraint
will become much more stringent after the completion of the E821 
experiment. If a discrepancy between the future E821 result and the SM 
prediction should emerge, models with a superlight gravitino might provide 
a viable explanation.

%
 
%

%
\section*{References}

\end{document}